# DeeptDCS: Deep Learning-Based Estimation of Currents Induced During Transcranial Direct Current Stimulation

Xiaofan Jia, *Student Member, IEEE*, Sadeed Bin Sayed, *Member, IEEE*, Nahian Ibn Hasan, *Student Member, IEEE*, Luis J. Gomez, *Member, IEEE*, Guang-Bin Huang, *Senior Member, IEEE*, and Abdulkadir C. Yucel, *Senior Member, IEEE*

*Abstract*— *Objective:* Transcranial direct current stimulation (tDCS) is a non-invasive brain stimulation technique used to generate conduction currents in the head and disrupt brain functions. To rapidly evaluate the tDCS-induced current density in near real-time, this paper proposes a deep learning-based emulator, named DeeptDCS. *Methods*: The emulator leverages Attention U-net taking the brain volume conductor models (VCMs) of head tissues as inputs and outputting the three-dimensional current density distribution across the entire head. The electrode configurations are also incorporated into VCMs without increasing the number of input channels; this enables the straightforward incorporation of the non-parametric features of electrodes (e.g., thickness, shape, size, and position) in the training and testing of the proposed emulator. *Results*: Attention U-net outperforms standard U-net and its other three variants (Residual U-net, Attention Residual U-net, and Multi-scale Residual U-net) in terms of accuracy. The generalization ability of DeeptDCS to non-trained electrode configurations can be greatly enhanced through fine-tuning the model. The computational time required by one emulation via DeeptDCS is a fraction of a second. *Conclusion*: DeeptDCS is at least two orders of magnitudes faster than a physics-based open-source simulator, while providing satisfactorily accurate results. *Significance*: The high computational efficiency permits the use of DeeptDCS in applications requiring its repetitive execution, such as uncertainty quantification and optimization studies of tDCS.

## I. INTRODUCTION

TRANSCRANIAL direct current stimulation (tDCS) is a non-invasive brain stimulation technique used to modify

Manuscript received April 22, 2022. This work was supported by Nanyang Technological University under a Start-Up Grant. (Corresponding author: *Abdulkadir C. Yucel.*)

X. Jia, S. B. Sayed, G.-B. Huang, and A. C. Yucel are with the School of Electrical and Electronic Engineering, Nanyang Technological University, Singapore 639798. (e-mails: xiaofan002@e.ntu.edu.sg, {sadeed.sayed, egbhuang, acyucel}@ntu.edu.sg).

N. I. Hasan and L. J. Gomez are with the Department of Electrical and Computer Engineering, Purdue University, West Lafayette, IN 47907, USA (e-mail: {hasan34, ljgomez}@purdue.edu).

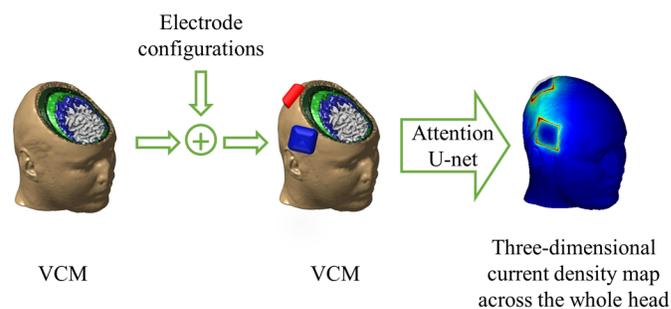

Electrode configurations

VCM　　　　　VCM　　　Attention U-net

Three-dimensional current density map across the whole head the activity of targeted brain regions. It has been widely used for research on cognitive enhancement [1]. Furthermore, its applications for treating the symptoms of various neurological and psychiatric disorders, including depression, epilepsy, schizophrenia, Alzheimer's disease, and chronic pain, have been investigated [2-6]. During tDCS, two (or more) electrodes are placed on the scalp and a weak current is injected through the electrodes. This current causes conduction currents that flow through the targeted regions of the brain. The precise biophysical mechanism of tDCS is still being investigated; however, it is hypothesized that conduction currents either facilitate or impede the excitability of targeted cortical neurons [7]. To maximize the current flowing through the targeted regions, electrodes are typically placed on opposite sides or surrounding areas of the targeted regions [8]. However, the optimum electrode placement, or montage, is subject-specific and it is recognized that knowledge of both the directions and intensities of currents in the brain is useful for optimizing the effects of tDCS [9]. To this end, this paper presents a deep learning-based (near) real-time emulator for determining the directions and intensities of the tDCS-induced current density inside the human head.

Fig. 1. The workflow of proposed DeeptDCS: a subject-specific VCM incorporating electrode information is input to Attention U-net imaging the three-dimensional current density across the whole head during tDCS.

Physics-based simulators have become the tool of choice for accurate modeling and determining tDCS-induced currents, or equivalently electric fields (E-fields). This is because measuring tDCS-induced fields in-vivo is limited to a few sparse regions and requires invasive surgical procedures [10], or is in experimental stage [11]. Furthermore, currents induced during tDCS are known to be sensitive to inter- and intra-subject variations [12, 13] and the position and shape of the electrodes [14]. Several simulators have been used for modeling tDCS-induced currents. Some of these simulators are freely distributed as software packages, such as SimNIBS [15] and ROAST [16], and are commonly used to inform about the practices of tDCS. These physics-based simulators require a long execution time involving around tens of minutes for head model generation and simulation, limiting their use in the applications where their repeated execution is required. These applications include individualized optimization of electrode shape and electrode placement over the scalp, and quantification of inter- and intra-subject uncertainty.

Recently, convolutional neural network (CNN) based tools have been developed for reducing the computational time required for head model generation from hours to seconds [17-19]. However, the use of CNN has not been investigated yet for reducing the computational time required for the simulation of tDCS. The use of CNNs for determining the currents can drastically reduce the computation time for simulation. The framework that can combine the CNNs for head model generation and simulation can proceed from magnetic resonance image (MRI) to currents in a few seconds. This, in turn, can enable its routine use for a broader set of tDCS applications, such as subject-specific optimization of electrode positions to target brain areas and quantification of uncertainties inherent in the tissues and tDCS setup.

CNN emulators for physics-based modeling have been explored in the context of research applications of transcranial magnetic stimulation (TMS), another non-invasive transcranial brain stimulation technique [20-22]. In these emulators, the computational burden is shifted to the offline training stage that is done only once, thereafter, fast and accurate emulations are performed online. However, the physical mechanisms of TMS and tDCS are different. The excitation for TMS is a primary volumetric E-field, which is different from the boundary electrode currents that drive tDCS. Moreover, in TMS, the E-fields can be partitioned into two components; one solely depends on the coil geometry and the other one depends on the head geometry. The component depending on the coil geometry (primary E-fields) accounts for the majority of the total E-field, while the one depending on the head geometry (secondary E-fields) accounts for around 30% of the total E-fields [23]. Deep learning-based TMS emulators learn the secondary E-fields while the primary E-field is calculated exactly. Furthermore, the primary E-field information is often provided as an input to the network [21] and some networks only predict the magnitude of the E-fields [20, 22]. As a result, the total E-field predicted by these approaches has a significantly lower error than the emulator-predicted secondary field. For tDCS, the fields are solely determined by the boundary currents flowing into the head from the boundary electrodes and are highly dependent on tissue geometry and conductivities [24]. To this end, the prediction of tDCS currents by deep learning is more challenging than that of TMS as the total field has to be computed by the emulator.

This study presents DeeptDCS, a deep learning-based emulator for rapidly determining currents induced in the head during tDCS. The proposed emulator leverages Attention U-net [25] as the deep learning technique. The Attention U-net was selected after its performance being compared with those of standard U-net [26], Residual U-net (Res U-net) [27], Attention Residual U-net (Attention Res U-net), and Multi-scale Residual U-net (MSResU-net) [21], which were previously employed in TMS emulations [20-22]. Attention U-net takes the volume conductor model (VCM) of head tissues and tDCS montages as input and outputs the three components of the tDCS-induced current density across the human head. Inclusion of the montage configuration in the VCM of head tissues allows keeping the input feature dimension as low as possible. The network is trained to minimize mean squared error loss with an input weighting scheme to suppress noise in the background. The well-trained network is evaluated on a testing dataset achieving a mean $\ell_2$-norm error ($M\ell_2E$) of 9.35% and a mean absolute error (MAE) of 0.000963 A/m$^2$, corresponding to 0.02476% of the maximum absolute value in the ground truth. Furthermore, the trained network is used to emulate tDCS-induced currents for non-trained montage positions, shape, and size (i.e., montage configurations not included in the training and testing datasets) after fine-tuning the model. In short, the contribution of this paper is threefold: (i) it serves as a "proof of concept" for the application of a deep learning algorithm to the tDCS emulation for the first time, (ii) it explains how to structure the inputs for the tDCS emulation via deep learning algorithm, and (iii) it compares the performances of existing deep learning techniques developed for TMS emulation in the past for the tDCS emulation.

The computational efficiency of DeeptDCS is demonstrated via comparison with a physics-based simulator, SimNIBS [15], an open-source package modeling brain stimulation techniques. Note that each simulation performed by SimNIBS requires 112 s on a CPU while SimNIBS cannot be executed on a GPU. Remarkably, each emulation performed by DeeptDCS takes 0.465 s on a GPU or 1 s on the same CPU. This makes the proposed emulator at least 112x faster than SimNIBS. Furthermore, when combined with CondNet [18], a CNN used to obtain the VCMs from subjects' MRIs in a few seconds, the proposed DeeptDCS allows near real-time visualization of current density distribution for a given MRI in clinical and research applications of tDCS. That said, the proposed DeeptDCS is intended to be used in applications requiring near real-time estimation of the current density with satisfactory accuracy. It should be noted that the accuracy of the results is satisfactory given the inherent uncertainties in the tDCS setup [12, 15, 28], which can be as large as 17%. It is the authors' opinion that current deep learning techniques cannot provide as accurate results as those provided by standard physics-based simulators. We believe that as new deep network structures and techniques are discovered, this opinion will change; the current study is the first systematic step toward this goal for the tDCS emulation.

Note that the proposed emulator DeeptDCS is described in detail for the first time in this paper, although a presentation outlining initial developments on DeeptDCS has been given at an earlier symposium [29]. In particular, the conference paper presented preliminary results obtained via standard U-net, while this paper provides a comprehensive performance comparison among five neural networks and concludes that Attention U-net achieves the highest accuracy among standard U-net and its variants. Moreover, architectures of employed neural networks, strategies for dataset generation, the input weighting scheme, and model fine-tuning for non-trained electrode positions are all elaborated in this paper but not in the conference paper.

## II. METHODOLOGY

Fig. 1 depicts the framework of DeeptDCS. DeeptDCS combines features for electrode configurations into a subject's VCM and deploys Attention U-net to emulate the tDCS-induced three-dimensional current density rapidly and accurately. In what follows, details of VCM generation are explained first. After the method incorporating montage features into VCM is discussed, the architectures of deep neural networks applicable to DeeptDCS are described. Finally, the loss function and input weighting scheme for training the networks are expounded.

### A. Generation of VCMs

The VCMs are generated using the *headreco* tool of SimNIBS [28]. SimNIBS leverages the finite element method to estimate induced fields in the full head model and uses other freely available software in the background such as FreeSurfer[30], FSL[31], and GMSH[32]. Utilizing the tetrahedral mesh, the software can be used for transcranial electrical and magnetic stimulation studies. The cortical segmentation tool *headreco* in SimNIBS can generate a virtual segmented head model consisting of different head tissue compartments. Constructed head models are co-registered based on EEG 10-10 electrode configuration [33] using a set of pre-defined fiducial points. The tool generates a tetrahedral finite-element model from two different MRI sequences namely, T1-weighted and T2-weighted sequences. Besides, the tool segments and assigns tissue conductivity from Atlas-based studies. In this study, six tissues are considered: white matter, grey matter, cerebrospinal fluid (CSF), skull, scalp, and vitreous humor. Furthermore, the uncertainty in tissue conductivity is also incorporated via Latin hypercube sampling. The tissue conductivities are selected from empirical studies in [12], shown in Table I.

**TABLE I**
**TISSUE CONDUCTIVITIES**

| Tissue | Notation | Min (S/m) | Max (S/m) | Standard (S/m) |
|--------|----------|-----------|-----------|----------------|
| White matter | $\sigma_1$ | 0.1 | 0.4 | 0.126 |
| Gray matter | $\sigma_2$ | 0.1 | 0.6 | 0.275 |
| CSF | $\sigma_3$ | 1.2 | 1.8 | 1.654 |
| Skull | $\sigma_4$ | 0.003 | 0.012 | 0.01 |
| Scalp | $\sigma_5$ | 0.2 | 0.5 | 0.465 |
| Vitreous humor | $\sigma_6$ | 0.3 | 0.7 | 0.5 |

Attention U-net operates on voxel meshes. To generate the voxel mesh of a head model, tetrahedral head model obtained from SimNIBS is converted into a voxel mesh in the dimension of $N \times N \times N$ to serve as the input of Attention U-net. To do that, the co-registered subject scalp is oriented along $x$, $y$, and $z$ directions from back to front (in sagittal plane), from left to right (in coronal plane) and from bottom to top (in transverse plane), respectively. First, the maximum axial dimension of each plane of the tetrahedral head model is calculated via $d_t = t_{max} - t_{min}$ , where $t \in \{x, y, z\}$ , $t_{max} = \max\{t_i; t_i \in \text{head}\}$ ,

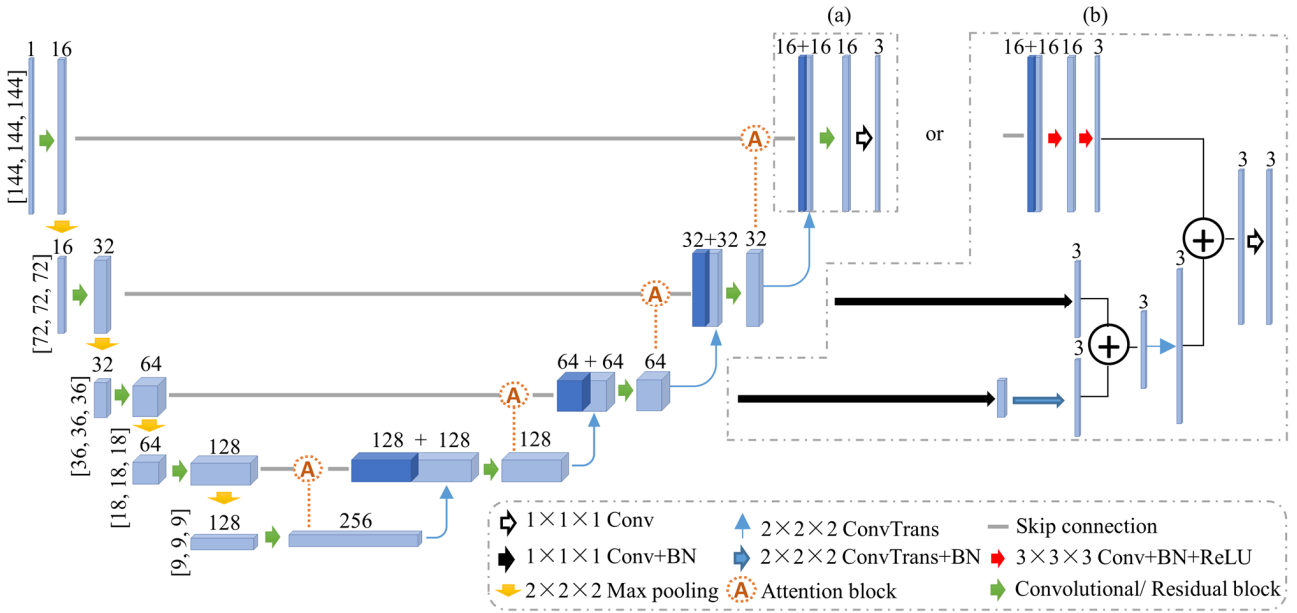

Fig. 2. Architectures of Attention U-net, standard 3D U-net, Res U-net, Attention Res U-net, and MSResU-net being compared for the proposed DeeptDCS emulator, where Conv and BN represent convolution and batch normalization, respectively. The attention block is only employed in Attention U-net and Attention Res U-net. (a) Output layers used by standard U-net, Res U-net, Attention U-net, and Attention Res U-net. (b) Output structures adopted by MSResU-net where features extracted from multi-level decoders are merged.

and $t_{\min} = \min\{t_i ; t_i \in \text{head}\}$ . The voxel edge length $\Delta d$ is determined by $\Delta d = \max\{d_i / N\}$ Then, the conductivity values of each tetrahedron center are mapped to the nearest voxels of the data grid by the *pointLocation* command in MATLAB. Voxels overlapped with non-head tissue regions are considered background (air) voxels. In this process, the number of voxels is not necessarily the same along each dimension. Hence, the grid is further padded with air voxels with zero conductivity until the grid size is $N \times N \times N$ . Padding is performed equally on both sides of the axis so that voxels occupied by head tissues and montages are centered in the grid. Moreover, the ground truth current density map is generated in the same fashion but consists of three channels corresponding to the $J_x$ , $J_y$ , and $J_z$ , thus, in size of $N \times N \times N \times 3$ .

### B. Incorporation of Electrode Features

The electrode placement significantly determines the magnitude and direction of the current in each head compartment [34]. To incorporate the electrode information, conductivity values of the voxels occupied by the electrodes are set slightly larger than those of the voxels corresponding to head tissues. Therefore, electrodes can be distinguished from head tissues in the inputs, thus guiding U-nets to pay more attention to features around the montages. This choice avoids the use of an additional input channel (and dimensionality), as the electrode information is incorporated into VCM directly. Furthermore, it enables non-parametric incorporation of electrode features, such as the thickness, shape, size, and positions of electrodes, to be captured by the emulator without significant additional efforts. In this study, tDCS montages consisting of two square saline layer electrodes attached to the scalp are considered.

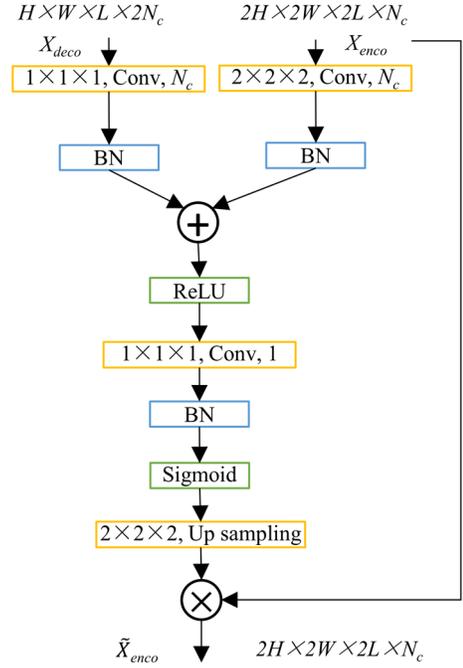

Fig. 4. Structure of attention block employed in Attention U-net and Attention Res U-net.

### C. 3D U-nets

To emulate the induced current density, DeeptDCS employs volume-to-volume a deep neural network, Attention U-net [25]. The attention unit incorporates additional connections to the standard U-net [26, 35] to better emphasize salient features in the encoding path. Standard U-net and its variants were first proposed for biomedical image segmentation and have been successfully deployed to predict E-fields induced in TMS [20-22].

Our implementation of the Attention U-net has a single channel input consisting of the VCM with the electrodes incorporated, while the three-channel output is the three-dimensional current density through the whole head, $J_x$ , $J_y$ , and $J_z$ . The details of the Attention U-net, standard U-net, and its other variants compared in this study are given below:

#### 1) Standard U-net

Fig. 2(a) (without attention blocks) illustrates the standard U-net architecture consisting of an encoder-decoder structure with five resolution steps and 3D operators. The contracting encoder path reduces spatial information and captures high-level features of VCMs while the successive expanding path reproduces the volume size and predicts current density distribution. Each step of the encoder section consists of three operations – convolution, batch normalization, and max-pooling [Fig. 3(a)]. The convolution operator uses $3 \times 3 \times 3$ kernels with a stride of $1 \times 1 \times 1$. Each convolution layer is followed by a batch-normalization (BN) layer and a max-pooling layer, respectively. The max-pooling layer consists of $2 \times 2 \times 2$ kernel. In the decoder, each layer contains a transposed convolution with a kernel size of 2 and stride length of 2 in every dimension, followed by the same convolution operator as in the encoding path. Rectified linear unit (ReLU) is applied wherever an activation function is required. The skip connection concatenates high-level features in the contracting path to the expanding path, merging local details with global information. The output convolutional layer with 1×1×1 kernel squeezes the number of channels into three, which corresponds to the predicted $J_x$, $J_y$, and $J_z$.

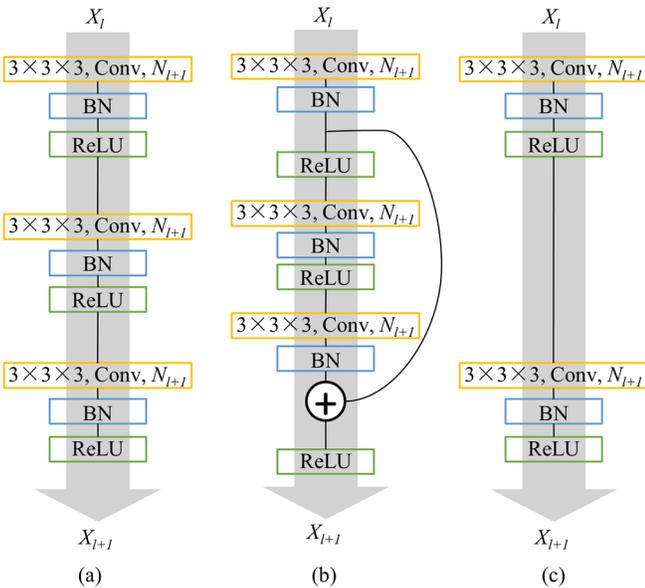

Fig. 3. (a) Convolutional block adopted by standard U-net and Attention U-net. (b) Residual block employed by Res U-net, Attention Res U-net, and encoding path of MSResU-net. (c) Convolutional block used in the decoding path of MSResU-net.

### 2) Attention U-net

Attention U-net has been adopted to estimate the magnitude of TMS-induced E-fields [22]. Attention U-net [25] boosts the performance of the standard U-net with attention blocks [Fig. 4], which takes encoding and decoding feature maps to generate a gating signal. Instead of direct concatenation via skip connections, the gating signal highlights salient features and suppresses noise in the feature maps that encoders extracted.

### 3) Other U-net Variants

Except for Attention U-net, standard U-net and its other variants have been used to emulate TMS-induced E-fields. Therefore, their capability for emulation of tDCS-induced currents is also investigated in this study.

**Res U-net** replaces convolutional blocks [Fig. 3(a)] in the standard U-net with residual blocks [Fig. 3(b)] [27]. With the identity mapping, residual blocks can resolve the gradient vanishing problem arising during the training of CNNs, thereby enabling the training of very deep networks. Further, neural networks with residual modules converge faster than their counterparts with convolutional blocks [27].

**Attention Res U-net** embeds both attention blocks [Fig. 4] and residual blocks [Fig. 3(b)] into the framework of standard U-net.

**MSResU-net** was developed in [36] for segmentation of MRIs and has been used to predict the E-fields induced during TMS [21]. Similar to the network developed in [21], the encoders of MSResU-net adopt residual blocks [Fig. 3(b)], while the decoders use convolutional blocks [Fig. 3(c)]. Furthermore, as demonstrated in Fig. 2(b), transposed convolutions in MSResU-net upscale high-level features to the dimensionality of outputs, therefore, three-level feature maps extracted by different level decoders are merged to predict the final current density distribution.

### D. Loss Function and Input Weighting Scheme

Attention U-net is trained to minimize a mean squared error (MSE) loss, which is

$$\text{MSE} = \frac{1}{N_{train} \times 3M} \sum_{s=1}^{N_{train}} \sum_{i=1}^{M} \sum_{j=1}^{3} (y_{i,j}^s - \tilde{y}_{i,j}^s)^2 \quad (1)$$

Here, $N_{train}$ is the number of training samples, $M$ is the number of voxels, while $y_{i,j}$ and $\tilde{y}_{i,j}$ represent the ground truth and predicted value for the $j^{\text{th}}$ component of current density in the $i^{\text{th}}$ voxel, respectively.

Moreover, an input weighting scheme filters the noise in the background of predictions. In the output of the network,

predicted values of air voxels are set to zeros since the indices of air voxels in the output are the same as those in the input.

## III. NUMERICAL RESULTS

This section explains the dataset used to train and test the DeeptDCS. It compares the performance of different neural networks applicable to DeeptDCS and presents the accuracy and computational efficiency of the Attention U-net embedded DeeptDCS. Finally, it explains how to generalize DeeptDCS for the non-trained electrode configurations.

### A. Dataset Generation

The dataset is constructed from 85 MRIs obtained from [37-40]. Classical tDCS montage configurations are adopted in this study. Two electrodes in size of 5×5 cm² and thickness of 5 mm are attached to the scalp; the currents injected through these electrodes are set to ±1 mA. Following recommendations in [34, 41], five commonly used montage configurations are modeled: CP5-CP4, TP7-Cz, F4-F3, CP5-Cz, and TP7-CP4 (anode-cathode) based on EEG 10-10 system.

For each montage configuration of a single subject's MRI, 200 VCMs with different tissue conductivity values are obtained via Latin hypercube sampling. Thereby, 85,000 samples are generated for five montages on 85 subjects' MRIs in total. The finite-element based tDCS solver (implemented in SimNIBS) is used for calculating the current density at each tetrahedral element of the VCMs.

Considering the computational resources and keeping high enough distinctive tissue-boundary resolution, a cubic grid size of 144×144×144 is chosen for the voxelized VCM and 144×144×144×3 for the current density map. Conductivities of anodes and cathodes are set to 2 S/m and 3 S/m in the voxelized VCMs, respectively, which are different from their real conductivities (29.4 S/m for silicone rubber and 1.0 S/m saline) used in SimNIBS for ground truth generation. As real conductivities of electrodes are over twenty times larger than head tissue conductivities, the adoption of the real electrode conductivities results in less distinctions between head tissues, which misleads U-nets and makes the optimization during training time-consuming.

### B. Training Strategies and Performances Comparison

The dataset of 85,000 samples (of 85 MRIs) is divided into three subsets following the standard 70%-15%-15% splitting: 59,000 samples (of 59 MRIs) for training, 13,000 samples (of 13 MRIs) for validation, and 13,000 samples (of 13 MRIs) for testing. The neural networks are implemented in Python using the deep learning platform of Keras [42] with TensorFlow backend [43]. Those are executed on a Linux server with an NVIDIA Tesla V100 GPU.

TABLE II
PERFORMANCE COMPARISON OF U-NETS

| | MAE ( A/m² ) | Mℓ₂E | RE | MRDM |
|---|---|---|---|---|
| Standard U-net | 0.001017 | 9.51% | 12.63% | 0.2101 |
| **Attention U-net** | **0.000963** | **9.35%** | **11.93%** | **0.1883** |
| Res U-net | 0.001028 | 9.78% | 12.75% | 0.2226 |
| Attention Res U-net | 0.001028 | 10.19% | 12.71% | 0.1969 |
| MSResU-net | 0.001124 | 10.30% | 13.99% | 0.2075 |

The network is trained using Adam optimizer [44] to minimize MSE between the ground truth current density and the current density estimated U-nets. The learning rate is initially set to 0.001 and subsequently divided by two if the validation loss does not decrease in the previous epoch. The batch size is set to three. Training continues until no apparent decrease in MSE of the validation set is observed and the best model with the smallest validation loss is saved for testing. To test the performance of trained networks in emulating the tDCS process, VCMs of 13 independent subjects (13,000 samples) not used in the training stage are considered.

The performance of the five models is compared by computing the mean absolute error (MAE), mean $\ell_2$-norm error ( $M\ell_2E$ ), relative error (RE) [45], and mean relative difference measure (MRDM) [46, 47] in the testing set, which are defined as

$$\text{MAE} = \frac{1}{N_{test} \times 3M} \sum_{s=1}^{N_{test}} \sum_{i=1}^{M} \sum_{j=1}^{3} \left| y_{i,j}^s - \tilde{y}_{i,j}^s \right| \tag{2}$$

$$\text{M}\ell_2\text{E} = \frac{1}{N_{test}} \sum_{s=1}^{N_{test}} \sqrt{\sum_{i=1}^{M} \left( \left\| \mathbf{y_i^s} \right\| - \left\| \mathbf{\tilde{y}_i^s} \right\| \right)^2 / \sum_{i=1}^{M} \left( \left\| \mathbf{y_i^s} \right\| \right)^2} \tag{3}$$

$$= \frac{1}{N_{test}} \sum_{s=1}^{N_{test}} \left( l_2 E \right)_s$$

$$\text{RE} = \frac{1}{N_{test}} \sum_{s=1}^{N_{test}} \left( \sum_{i=1}^{M} \left\| \mathbf{y_i^s} - \mathbf{\tilde{y}_i^s} \right\| / \sum_{i=1}^{M} \left\| \mathbf{y_i^s} \right\| \right) \tag{4}$$

$$\text{MRDM} = \frac{1}{N_{test}} \sum_{s=1}^{N_{test}} \left( \frac{1}{M^s} \sum_{i=1}^{M^s} \left\| \frac{\mathbf{y_i^s}}{\left\| \mathbf{y_i^s} \right\|} - \frac{\mathbf{\tilde{y}_i^s}}{\left\| \mathbf{\tilde{y}_i^s} \right\|} \right\| \right). \tag{5}$$

$N_{test}$ is the number of testing samples, $\left\| \mathbf{y_i^s} \right\|$ and $\left\| \mathbf{\tilde{y}_i^s} \right\|$ are the magnitudes of ground truth and predicted current density of $i^{th}$ voxel, respectively, and $\left\| \mathbf{y_i^s} \right\| = ((y_{i,1}^s)^2 + (y_{i,2}^s)^2 + (y_{i,3}^s)^2)^{0.5}$, $\left\| \mathbf{\tilde{y}_i^s} \right\| = ((\tilde{y}_{i,1}^s)^2 + (\tilde{y}_{i,2}^s)^2 + (\tilde{y}_{i,3}^s)^2)^{0.5}$. $l_2 E$ stands for the $\ell_2$-norm error of the $s^{th}$ testing sample. $M^s$ denotes the number of voxels occupied by the head model in the $s^{th}$ testing sample. MRDM evaluates the angular accuracy of the predictions and should be in the range [0, 2]. Table II shows the obtained error rates of the five CNNs. The lowest error rates are achieved by Attention U-net, therefore it is chosen to be used in DeeptDCS. Training Attention U-net takes eight days. The MSE loss of Attention U-net for training, validation, and testing are $1.5276 \times 10^{-5}$, $2.0388 \times 10^{-5}$, and $1.9840 \times 10^{-5}$, respectively.

### C. TDCS Emulations by Attention U-net

#### 1) Performance on the testing set

Fig. 5 illustrates the current density computed midway between the gray and white matter compartment of three testing samples. The predicted voxel current-density magnitudes (the second column) are compared with the ground-truth distribution (the first column) based on three electrode montages, namely, (a) TP7-Cz, (b) F4-F3, and (c) CP5-CP4. As shown in Fig. 5, the DeeptDCS can predict current density in both high and low excitation regions of the brain with minimal error. The high consistency of the current density in the first two columns and the small absolute difference (in the third column) between them confirms the accuracy of DeeptDCS and its ability to localize highly excited brain regions during tDCS. Relative differences ( the fourth column) are normalized to the maximum ground truth current density in the brain region, which is defined as $\left\| \mathbf{y_k^s} \right\| - \left\| \mathbf{\tilde{y}_k^s} \right\| / \max_k \left\| \mathbf{y_k^s} \right\|$, $\mathbf{k} \in$ brain. Although a few red spots can be observed in figures of relative difference, the values of other voxels are small demonstrating a good agreement between the prediction from DeeptDCS and the ground truth from SimNIBS.

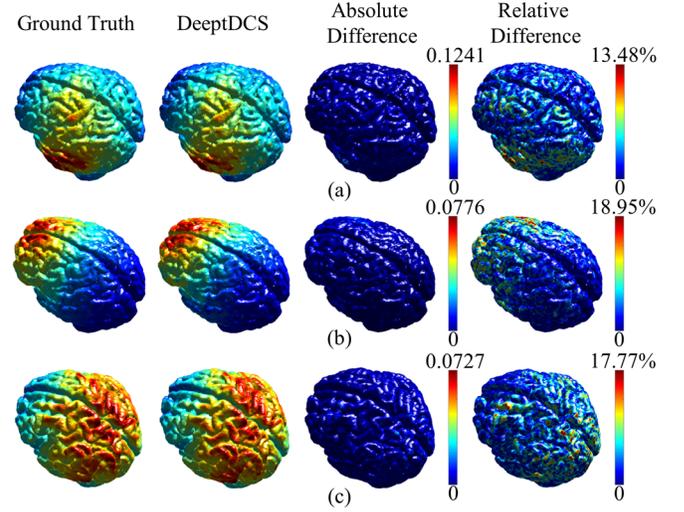

Fig. 5. 3D visual comparison of magnitudes of current density across the brain, where the four columns represent the ground truth from SimNIBS, the emulation from proposed DeeptDCS, the corresponding absolute difference, and relative difference. Square electrodes in size of 5×5cm² are placed at (anode-cathode) (a) TP7-Cz, (b) F4-F3, and (c) CP5-CP4 based on EEG 10-10 system. Tissue conductivities ( $\sigma_1$ to $\sigma_6$ ) are (a) 0.3048, 0.5160, 1.6259, 0.0084, 0.4387, 0.4211 S/m, (b) 0.3384, 0.4934, 1.5425, 0.0107, 0.4156, 0.3955 S/m, and (c) 0.2231, 0.4300, 1.3879, 0.0070, 0.3030, 0.6985 S/m, respectively. The unit of values in the figure is A/m² .

Figs. 6(a) and 6(b) compare the magnitudes of the predicted current density with those of the ground truth on 2D slices containing six tissues for two randomly chosen subjects from the testing set. The voxel-level agreement between the ground truth and the DeeptDCS output demonstrates the accuracy of the DeeptDCS for the studies involving the stimulation of deep brain compartments. Furthermore, DeeptDCS can also predict the direction of current density, $J_x$ , $J_y$ , and $J_z$ . Fig. 7 compares the directional flow of current obtained by the finite-element based simulator in SimNIBS (ground-truth) and predicted by the DeeptDCS. The directions and magnitudes of the current density vectors predicted by DeeptDCS match well with those obtained by SimNIBS.

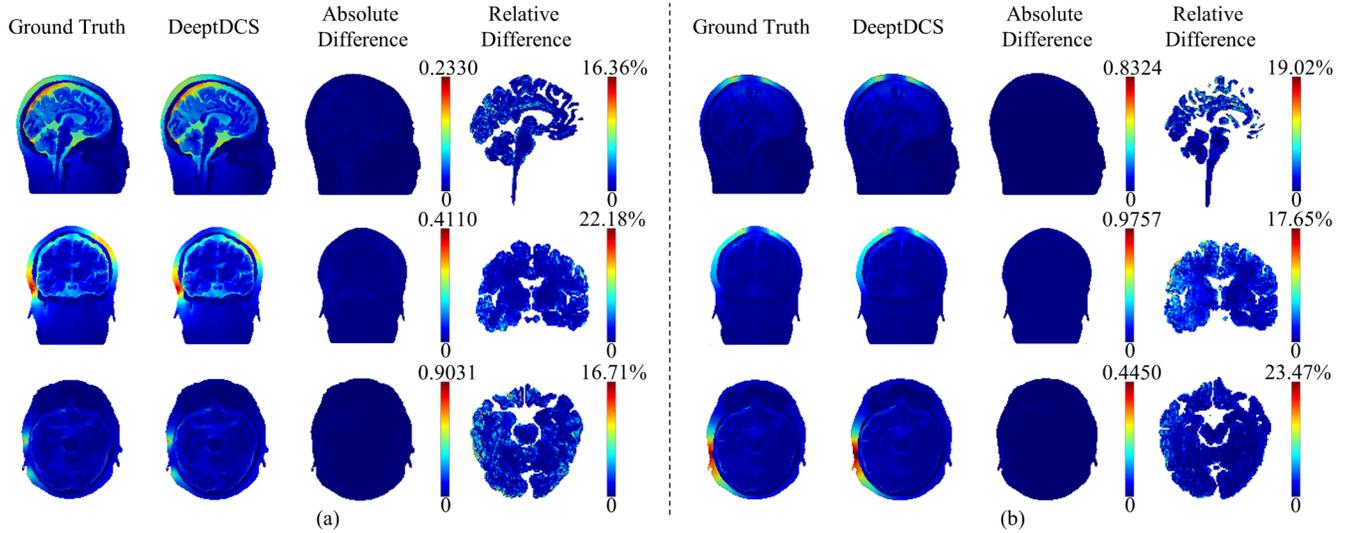

Fig. 6. The magnitudes of current density across the head for two test samples, (a) and (b), displayed in sagittal (the first row), coronal (the second row), and transverse (the third row) slices. Square electrodes in size of $5 \times 5 \text{cm}^2$ are placed at (a) TP7-CP4 and (b) CP5-Cz based on EEG 10-10 system. Tissue conductivities ($\sigma_1$ to $\sigma_6$) are (a) 0.1173, 0.4891, 1.3229, 0.0117, 0.2395, 0.3505 S/m and (b) 0.2107, 0.3588, 1.2063, 0.0115, 0.4586, 0.4976 S/m, respectively. The unit of values in the figure is A/m$^2$.

By considering all voxels in the computational domain, the MAE is computed as 0.000963 A/m$^2$, corresponding to 0.02476% of the maximum absolute value in the ground truth, while the M$\ell_2$E is obtained as 9.35%. M$\ell_2$E of six head tissues are 8.81% (white matter), 12.79% (gray matter), 10.38% (CSF), 18.00% (skull), 9.09% (scalp), and 9.22% (vitreous humor), respectively. The $\ell_2$-norm error ($\ell_2$E) distribution of all testing samples is illustrated in Fig. 8. 85% of the samples have less than 10.69% error. Obtained RE and MRDM are 11.93% and 0.1883, respectively.

To compare the efficiency of SimNIBS and DeeptDCS, a set of VCMs is constructed from four subjects' MRIs and five electrode positions used in the testing set. For each montage configuration of a single subject's MRI, ten VCMs with randomly selected tissue conductivity values are obtained. Thereby, 200 VCMs are generated and further used in SimNIBS and DeeptDCS separately to predict current density distributions. Time consumed by the 200 executions is averaged to obtain the required time for each estimation. The computational times of DeeptDCS and the physics-based simulator SimNIBS for a single emulation and simulation are 1 s and 112 s, respectively, on an Intel Xeon Gold 6128 CPU 3.40 GHz. However, when accelerated by an NVIDIA Tesla P100

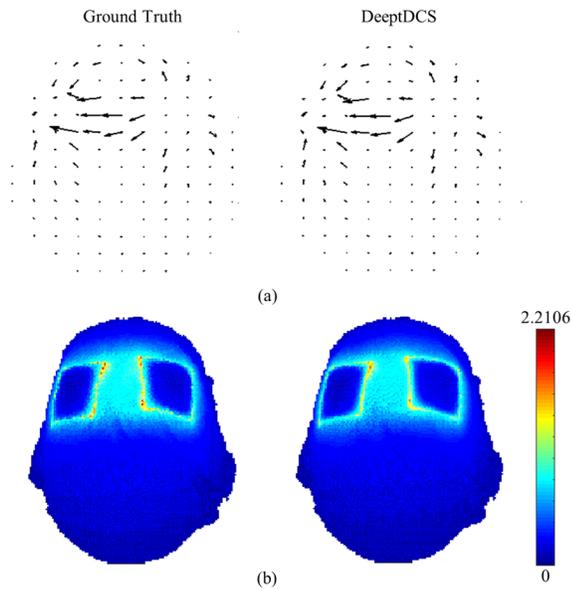

Fig. 7. Top view of the (a) directions and (b) magnitudes of volume current density. A perfect match between the ground truth and DeeptDCS-predicted current densities is observed. The unit of values in the figure is A/m$^2$.

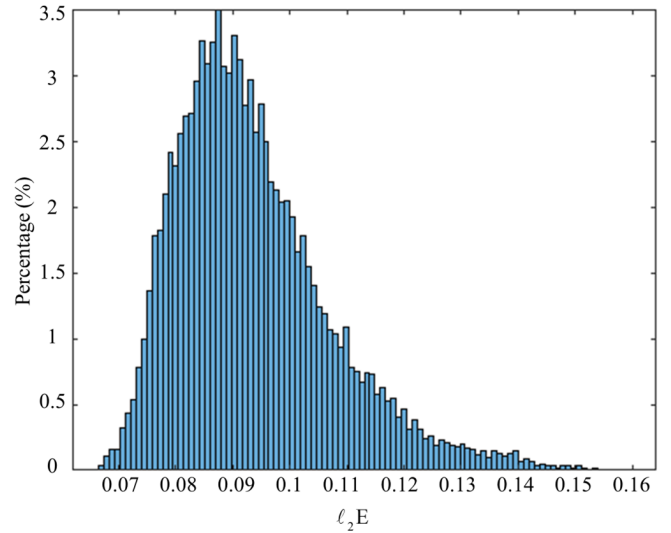

Fig. 8. $\ell_2$E distribution of the 13,000 testing samples constructed from 13 subjects' MRIs and five pre-trained electrode configurations.

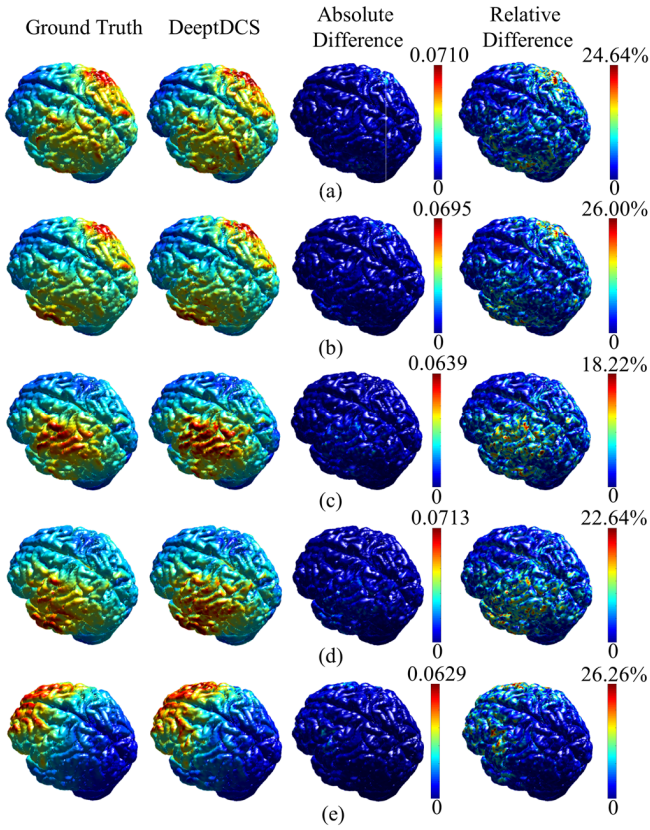

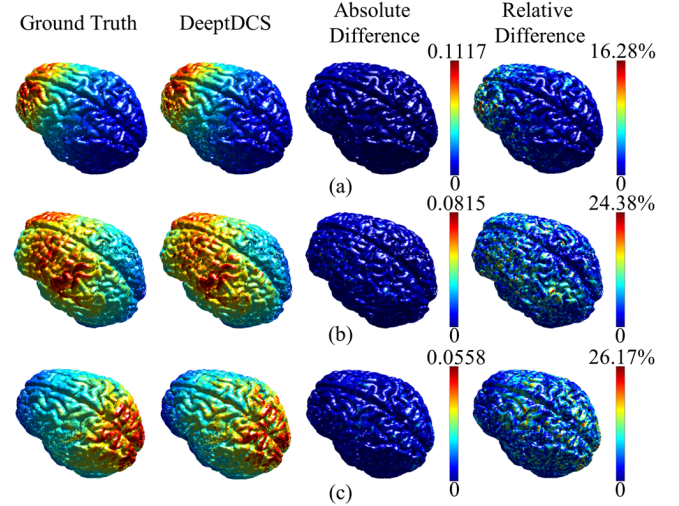

Fig. 9. Validation test results on standard tissue conductivity values listed in Table I. The magnitudes of current density across the brain are displayed. Four columns represent the ground truth from SimNIBS, the emulation from proposed DeeptDCS, the corresponding absolute difference and relative difference between results. Square electrodes in size of $5\times5cm^2$ are placed at (a) CP5-CP4, (b) TP7-CP4, (c) CP5-Cz, (d) TP7-Cz, and (e) F4-F3 based on EEG 10-10 system. The unit of values in the figure is $A/m^2$.

Fig. 10. 3D visual comparison of ground truth (from SimNIBS) and predictions (from DeeptDCS) on the testing set constructed from three **non-trained electrode positions**. *Square* electrodes in size of $5\times5cm^2$ are placed at (a) F4-Fp1, (b) C3-Fp2, and (c) Cz-POz based on EEG 10-10 system. Tissue conductivities ($\sigma_1$ to $\sigma_6$) are (a) 0.2363, 0.5920, 1.5984, 0.0084, 0.3831, 0.6744 S/m, (b) 0.2231, 0.4300, 1.3879, 0.0070, 0.3030, 0.6985 S/m, and (c) 0.2135, 0.3531, 1.7962, 0.0106, 0.3985, 0.6825 S/m, respectively. The unit of values in the figure is $A/m^2$.

GPU, DeeptDCS can perform one emulation in 0.456s with the testing batch size of one. Therefore, DeeptDCS executes at least 112x faster than the physics-based simulator while providing satisfactory accuracy. Apparently, DeeptDCS is suitable for near real-time applications requiring visualization of the tDCS-induced fields.

### 2) Validation test on standard tissue conductivity values

To evaluate the performance of DeeptDCS, five VCMs are generated from a new subject's MRI [48] (not utilized in training/validation/testing set) using the standard conductivity values listed in Table I. The proposed DeeptDCS is used to predict the current distributions of these VCMs for five montages and achieves MAE, $M\ell_2E$, RE, and MRDM of 0.001480 $A/m^2$ (0.06651% of the maximum absolute value in the ground truth), 10.91%, 16.26% and 0.2301. Fig. 9 shows the high agreement between the current distributions obtained by SimNIBS and the proposed DeeptDCS emulator.

### 3) Validation test on randomly selected conductivity values

DeeptDCS are expected to cope with variations in tissue conductivity values. Therefore, a new testing dataset including

100 samples is constructed from one MRI in the training set with electrodes placed at TP7-CP4 (anode-cathode). Tissue conductivity values are randomly selected from the reference range [Table I] via the Monte Carlo method. Pre-trained DeeptDCS obtains MAE, $M\ell_2E$, RE, and MRDM of 0.000853 $A/m^2$ (0.05265% of the maximum absolute value in the ground truth), 8.23%, 11.05%, and 0.1744, respectively, on the new dataset. These error rates are comparable to the previous testing results on the 85 MRIs dataset, proving that DeeptDCS can perform well on non-trained tissue conductivity values.

### D. Fine-tuning for Non-trained Electrode Configurations

Although trained on $5\times5 cm^2$ square montages placed at five positions, DeeptDCS can be deployed to emulate tDCS under non-trained electrode positions, shape, and size via fine-tuning the model (also known as transfer learning). To do that, a base model pre-trained on a large and similar dataset can serve as a starting point to capture generic feature maps. Rather than training the CNN from scratch, fine-tuning a pre-trained base model on a smaller relevant dataset can efficiently improve the model performance on the new task. Fine-tuning of Attention U-net in DeeptDCS is executed on a Linux server with an NVIDIA Tesla P100 GPU with a batch size of one.

Three datasets I, II, and III are constructed from nine new subjects' MRIs (not included in previous 85 MRIs dataset). Electrode thickness is fixed at 5 mm. In a similar fashion, 200 VCMs with different tissue conductivity values (following the statistics in Table I) are obtained for each montage of a single subject's MRI via Latin hypercube sampling. In each new dataset, samples generated from seven subjects' MRIs are used

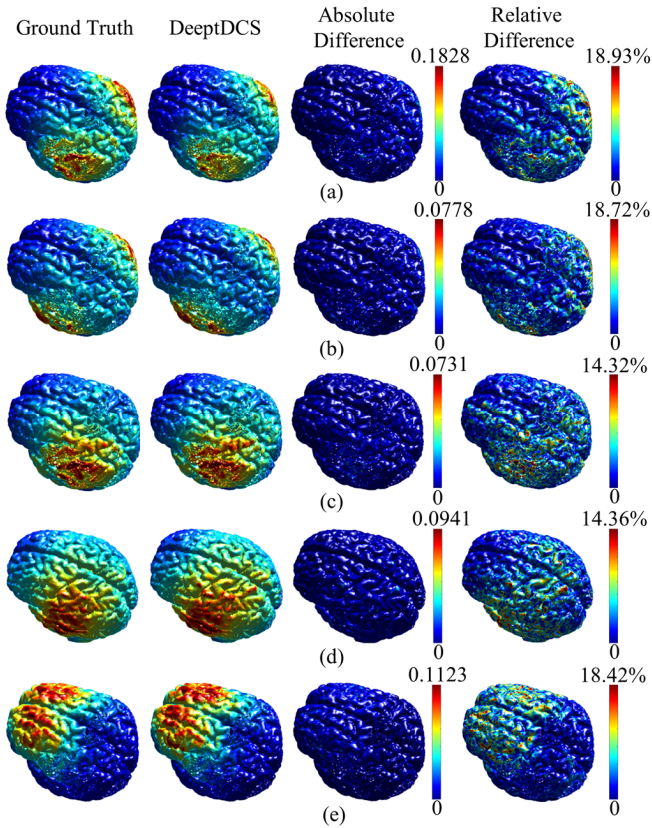

Fig. 11. 3D visual comparison of ground truth (from SimNIBS) and predictions (from DeeptDCS) on the testing set constructed from **non-trained electrode shape**. *Circular* electrodes with diameter of 5 cm are placed at (a) CP5-CP4, (b) TP7-CP4, (c) CP5-Cz, (d) TP7-Cz, and (e) F4-F3 based on EEG 10-10 system. Tissue conductivities ( $\sigma_1$ to $\sigma_6$ ) are (a) 0.2700, 0.3716, 1.4208, 0.0091, 0.2445, 0.5254 S/m, (b) 0.3401, 0.1909, 1.2100, 0.0046, 0.4116, 0.6600 S/m, (c) 0.2337, 0.2895, 1.2911, 0.0057, 0.4867, 0.6259 S/m, (d) 0.2231, 0.4300, 1.3879, 0.0070, 0.3030, 0.6985 S/m, and (e) 0.1863, 0.3771, 1.4769, 0.0096, 0.3410, 0.3477 S/m respectively. The unit of values in the figure is A/m$^2$.

for training, while the rest (from two subjects' MRIs) are used for testing.

### 1) Fine-tuning for non-trained electrode positions

Samples in dataset I are stimulated by 5×5 cm$^2$ square electrodes placed at three new positions, namely F4-Fp1 [49], C3-Fp2 [50], and Cz-Poz [51] (anode-cathode) based on EEG 10-10 system. In total, 5,400 samples are generated for three non-trained montages on nine subjects' MRIs.

Without any fine-tuning on the network parameters, the Attention U-net model pre-trained on 59,000 samples and five electrode configurations obtains an MAE of 0.006858 A/m$^2$ (0.31995% of the maximum absolute value in the ground truth), M$\ell_2$E of 52.40%, RE of 85.60%, and MRDM of 1.0817 on the new 4,200 samples dataset with three new electrode positions. However, by taking the pre-trained Attention U-net model as the starting point, training on the 4,200 samples reduces MAE to 0.001504 A/m$^2$ (0.07016% of the maximum absolute value in the ground truth), RE to 18.67%, and MRDM to 0.2595. Meanwhile, the fine-tuning yields a dramatic reduction of the testing M$\ell_2$E from 52.40% to 13.43%.

Fine-tuning on dataset I takes four hours. Fig. 10 provides visual comparison between the ground truth and the prediction on three representative testing samples, which demonstrates that DeeptDCS predictions are qualitatively satisfactory.

### 2) Fine-tuning for non-trained electrode shape

Dataset II uses *circular* electrodes with a diameter of 5 cm. Five electrode positions are the same as those used in the 85 MRIs dataset. Therefore, dataset II contains 9,000 samples. Fine-tuning the Attention U-net pre-trained on 85 MRIs dataset decreases MAE, M$\ell_2$E, and RE from 0.001833 A/m$^2$ (0.06147% of the maximum absolute value in the ground truth), 21.81%, and 20.42% to 0.001403 A/m$^2$ (0.04708% of the maximum absolute value in the ground truth), 11.36%, and 15.51%, respectively. Although MRDM slightly increases from 0.2301 to 0.2632, MRDMs obtained by the pre-trained model and the fine-tunned model are both acceptable. Fine-tuning on dataset II costs 21.9 hours. Fig. 11 demonstrates that predictions from DeeptDCS agree well with reference results obtained from SimNIBS.

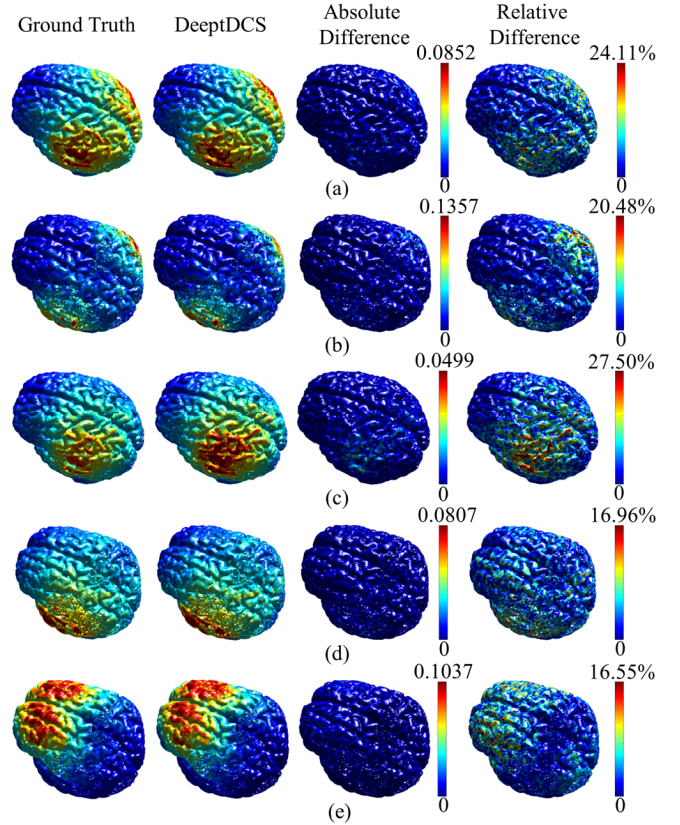

Fig. 12. 3D visual comparison of ground truth (from SimNIBS) and predictions (from DeeptDCS) on the testing set constructed from **non-trained electrode size**. *Square* electrodes in size of 4×4 cm$^2$ are placed at (a) CP5-CP4, (b) TP7-CP4, (c) CP5-Cz, (d) TP7-Cz, and (e) F4-F3 based on EEG 10-10 system. Tissue conductivities ( $\sigma_1$ to $\sigma_6$ ) are (a) 0.3348, 0.3028, 1.7904, 0.0067, 0.2171, 0.6579 S/m, (b) 0.3196, 0.1203, 1.3156, 0.0079, 0.2090, 0.5162 S/m, (c) 0.34401, 0.2248, 1.2141, 0.00623, 0.4896, 0.4830 S/m, (d) 0.2337, 0.2895, 1.2911, 0.0057, 0.4867, 0.6259 S/m, and (e) 0.2231, 0.4300, 1.3879, 0.0070, 0.3030, 0.6985 S/m respectively. The unit of values in the figure is A/m$^2$.

### 3) Fine-tuning for non-trained electrode size

Electrodes in dataset III are squares with size of $4\times4$ cm$^2$. Electrodes are placed at the same positions used in the 85 MRIs dataset. Totally, 9,000 samples are generated for dataset III. The direct application of Attention U-net pre-trained on 85 MRIs dataset yields relatively high error rates on the testing set of dataset III -- MAE of 0.002199 A/m$^2$ (0.07412% of the maximum absolute value in the ground truth), M$\ell_2$E of 27.66%, and RE of 24.35%. Nevertheless, parameter fine-tuning reduces MAE to 0.001399 A/m$^2$ (0.04716% of the maximum absolute value in the ground truth), M$\ell_2$E to 11.67%, and RE to 15.51%. MRDM increases from 0.2346 to 0.2562 after fine-tuning, however, both MRDMs are reasonable. Fine-tuning on dataset III requires 30 hours. Fig. 12 shows the match between the ground truth and DeeptDCS-emulated current density distribution.

## IV. DISCUSSION

U-net parameters have been optimized for the prediction of tDCS-induced current density map. The Attention U-net obtains the highest emulation accuracy on the testing set, as demonstrated in Table II. The hypothesized reason that the Attention U-net performs slightly better than the standard U-net is that attention gates highlight significant information in the encoding feature maps before they are concatenated to the decoder path. Regions with a high current density distribution and their magnitude can be identified using DeeptDCS results. Gating signal in the Attention U-net guides the network to put more attention to the feature corresponding to these areas, thereby resulting in a better performance. However, improvements observed relative to a standard U-net are marginal. We hypothesize this is because attention gates are also developed to suppress noise in the feature map. Nevertheless, the input weighting scheme in DeeptDCS fully removes the noise in the background by setting the value of air voxels to zeros, which disables the noise filtering contribution from attention blocks limiting their applicability.

Residual blocks and multi-level deep supervision in Res U-net, Attention Res U-net, and MSResU-net are not conducive to the model accuracy. This is because these two schemes are primarily proposed to address the gradient vanishing problem potentially existing in the training of very deep neural networks [27, 36]. Therefore, these two schemes may achieve higher accuracy if the backbone network faces a degradation problem, which is not observed here. The TMS emulator developed in [21] leveraged MSResU-net, however, since its performance was not compared with the standard U-net, it is not known whether this is important in that context.

DeeptDCS achieves an M$\ell_2$E of 9.35%, which is higher than the error introduces by many physics-based tDCS simulators. DeeptDCS is not proposed to substitute the physics-based solver but mainly for applications requiring near real-time emulation with satisfactory accuracy, such as optimal electrode placement during neuronavigation and uncertainty quantification studies. Furthermore, the error around 10% for all the five CNNs cannot solely be attributed to the CNNs' accuracy as it can also stem from the systematic errors, for instance, the error introduced by MRI segmentation and the uncertainties in head tissue conductivities [12, 28, 52].

## V. CONCLUSION

In this paper, a deep learning-based emulator, called DeeptDCS, was proposed for estimating the tDCS-induced current density. The emulator takes an individual's VCM with features of electrodes as the input and outputs the three components of the current density across the whole head via Attention U-net, which outperforms standard U-net and its other variants previously employed in TMS emulations. The visual comparisons between the ground truth and predicted results demonstrate the accuracy of DeeptDCS predictions. Furthermore, DeeptDCS can emulate induced currents under non-trained electrode configurations via fine-tuning the network. Notably, DeeptDCS is faster than the physics-based simulator, making it a strong candidate for the applications requiring repetitive testing, such as subject-specific optimization of electrode positions and uncertainty quantification studies.